\pdfoutput=1 
\documentclass[11pt]{article}
\usepackage{amssymb}
\usepackage{graphicx}
\parskip \baselineskip
\newcommand{\ds}{\displaystyle}

\newcommand{\Z}{\mathbb{Z}}

\newcommand{\ol}{\overline}

\newcommand{\ra}{\rightarrow}

\newcommand{\Mod}{\mbox{Mod}}

\title{Compression with wildcards: \\
All  $k$-models of a Binary Decision Diagram}

\author{Marcel Wild and Yves Semegni}
\date{}

\begin{document}
\maketitle

\begin{quote}
A{\scriptsize BSTRACT}: {\footnotesize Given a Binary Decision Diagram $B$ of a Boolean function $\varphi$ in $n$ variables, it is well known that all $\varphi$-models can be enumerated in output polynomial time, and in a compressed way (using don't-care symbols). We show that  all $N$ many $\varphi$-models of fixed Hamming-weight $k$ can be enumerated in time polynomial in $n$ and $|B|$ and $N$ as well. Furthermore, using novel wildcards, again enables a compressed enumeration of these models.}
\end{quote}

\section{Introduction}

Before we come to the Section break-up in 1.1, and to previous work relating to our topic in 1.2, let us recall two notions of enumeration. Often the desired objects (e.g. all spanning trees of a graph) are enumerated one-by-one but with {\it polynomial delay}. This means there is a polynomial $p(I)$ in the {\it input size} $I$ such that it takes time $\le p(I)$ until the first object is generated, and also time $\le p(I)$ between any two objects. An enumeration algorithm is said to run in {\it polynomial total time} if all instances run\footnote{ For the sake of clarity we assume that each output object is `small', i.e. of size $\le poly(I)$. Also note that instead of polynomial total time some authors speak of {\it output polynomial time} but this is unprecise since $I$ also enters $q$.} in time at most $q(I,N)$, where $q(I,N)$ is a polynomial in $I$ and the number $N$ of objects.  Obviously polynomial delay implies polynomial total time (put $q(I,N):=Np(I)$), but not conversely. In particular, `polynomial delay' does not make sense for algorithms that do not generate the objects one-by-one, but in some compressed format. A case in point is the algorithm in Section 4 which runs in polynomial total time $q(I,N)$. Here $I=|B|+n$ where $|B|$ is the size of a given binary decision diagram $B$ of a Boolean function $\varphi:\{0,1\}^n\to\{0,1\}$, and $N$ is the number of $\varphi$-models of Hamming-weight $k$.
Previous versions of the present article (the first from 2017) can be found in the arXiv.

{\bf 1.1} In the  sequel we assume a basic familiarity with Boolean functions [CH] and with Binary Decision Diagrams (BDD's), as e.g. provided in [K]. We adopt the nowadays common practise that BDD always means ordered BDD.
In Section 2 we review the standard method for calculating the {\it cardinality} of the model set Mod$(\varphi)$ from a BDD of a Boolean function $\varphi:\{0,1\}^n\ra\{0,1\}$. The {\it Hamming-weight} $H(x)$ of a bitstring $x\in\{0,1\}^n$ is the number of indices $i$ with $x_i=1$. Everything in this article centers around\footnote{Everything in our article most likely generalizes to models of fixed $f$-weight for any weight function $f:\{1,\ldots,n\}\to \Z$. However, restriction to the Hamming weight benefits the flow of ideas.}.

$$Mod(\varphi, k): = \{x \in \mbox{Mod}(\varphi): H(x) =k\}.$$

We henceforth refer to to the bitstrings in $Mod(\varphi, k)$ as {\it k-models}.
We follow Knuth [K] to find all cardinalities $|\mbox{Mod}(\varphi, k)|$  as the coefficients of a polynomial which can be calculated fast recursively. 
Section 3 reviews how a BDD of $\varphi$ allows $\Mod(\varphi)$ to be enumerated in a compressed fashion, using the don't-care symbol $*$.
In contrast, it is evident that the symbol $*$ cannot be used to compress $\Mod(\varphi,k)$; any partial model using $*$, say $(1,0,*,1)$, necessarily comprises $\varphi$-models of {\it different} Hamming weight, i.e. $(1,0,0,1)$ and $(1,0,1,1)$. 

This is cured in Section 4 by virtue of the wildcard
 $g_t g_t \cdots g_t$ which means ``exactly $t$ many 1-bits''. Thus the $01${\it g-row}
 $(g_2, 0,1, g_2, g_2)$ is the set $\{({\bf 1}, 0, 1, {\bf 1}, {\bf 0}), ({\bf 1}, 0, 1, {\bf 0}, {\bf 1}), ({\bf 0},0,1, {\bf 1}, {\bf 1})\}$. In practise even a naive application (Subsection 4.1) of 01g-rows for enumerating $\Mod(\varphi,k)$ can offer a dramatic improvement. Take for example the Boolean function $\varphi:\{0,1\}^{100}\ra\{0,1\}$ whose BDD consists simply of the root $x_1$, the $0$-son  $\perp$,  and the $1$-son $\top$. Then $|Mod(\varphi)|=|(1,*,...,*)|=2^{99}$. 
 Outputting the ${99}\choose{49}$ bitstrings in
$\Mod(\varphi,50)$ one-by-one is ridiculously infeasible, whereas the compressed representation boils down to $(1,g_{49},g_{49},\ldots,g_{49})$.
However, the worst cases of the naive method cannot be forced to run in polynomial total time. Fortunately this is mended in 4.2 (by Example) and 4.3 (the main Theorem).  Parts of the proof will be postponed to Section 5. An upper bound on the number of produced 01g-rows can be given {\it beforehand}. It is the sharper the smaller the height of the BDD (4.4).

In Section 6 we present numerical results obtained by the second author. For instance, some Boolean function $\varphi:\{0,1\}^{119}\to\{0,1\}$ in CNF format had about $10^{34}$ many 65-models. Representing $Mod(\varphi,65)$ as a disjoint union of $01g$-rows in the naive way required 113477 $01g$-rows and took 224 seconds. The sophisticated way used 964 rows and took 8 seconds.

As to applications, e.g. suppose Mod$(\varphi)$, provided by a BDD, is the family of all hitting sets of a set system. Often only small hitting sets are sought. The smallest $k$ for which Mod$(\varphi,k)$ is non-empty can be determined fast {\it beforehand} using Knuth's  method (Section 2). The algorithm of Section 4 then renders  Mod$(\varphi, k)$ as a disjoint union of few 01g-rows. 

{\bf 1.2} As to related research, the closest match seems to be the article [ABJM]. It deals with d-DNNF's, which is a format of Boolean functions $\varphi$ (due to Darwiche 2002) that properly subsumes BDD's. Let a d-DNNF of $\varphi$ be given. It is shown in [ABJM] that after linear preprocessing time all $k$-models of $\varphi$ can be generated with constant (thus polynomial) delay. While d-DNNF's are more general than BDD's, observe that the enumeration described in the lengthy article [ABJM] is one-by-one, and thus infeasible when $N$ is large. The example in 1.1 with 
$N={{99}\choose{49}}$ illustrates that point.

\section{Calculating $|\Mod(\varphi)|$ and $|\mbox{Mod}(\varphi,k)|$ }

Consider the BDD $B_1$ in Figure 1 which defines a unique Boolean function $\psi: \{0,1\}^{10} \ra \{0,1\}$. All nodes other than $\perp$ and $\top$ are called {\it branching nodes}. In particular the top node $x_1$ is the {\it root} of the BDD. (For the time being ignore all labels like $[3,4]$.) Recall how one decides whether or not a bitstring like
$$x = ({\bf x}_1, x_2, x_3, {\bf x}_4, x_5, x_6, {\bf x}_7, x_8, x_9, x_{10}) : = ({\bf 1},1, 0, {\bf 1}, 1, 0,  {\bf 0},1,1,0)$$
belongs to Mod$(\psi$). Start at the root $x_1$ in Figure 1. Because in our bitstring $x_1 =1$, go down on the solid line to the {\it $1$-son} $x_4$. From there, because our $x_4 =1$, branch again on the solid line and visit the  $1$-son $x_7$. Finally, in view of $x_7 =0$, go down the dashed line to the {\it $0$-son} $\perp$, which signals that $x \not\in \mbox{Mod}(\psi)$. In all of this the values $x_i$ for $i \not\in \{1, 4,7\}$ were irrelevant. Notice that speaking of `the' node $x_7$ is ambiguous because there are two nodes labeled $x_7$ in Figure 1. Hence, as in Figure 2, we need  distinct labels for distinct nodes.

For many purposes  a {\it shelling (from below)} is useful, i.e. one keeps  pruning, in any order, the minimal branching nodes of the shrinking BDD until it is exhausted. For instance $a, b, c, d, e, f$ in Figure 2,  but also $b, c, d, a, e, f$ are shellings.

\includegraphics[scale=0.45]{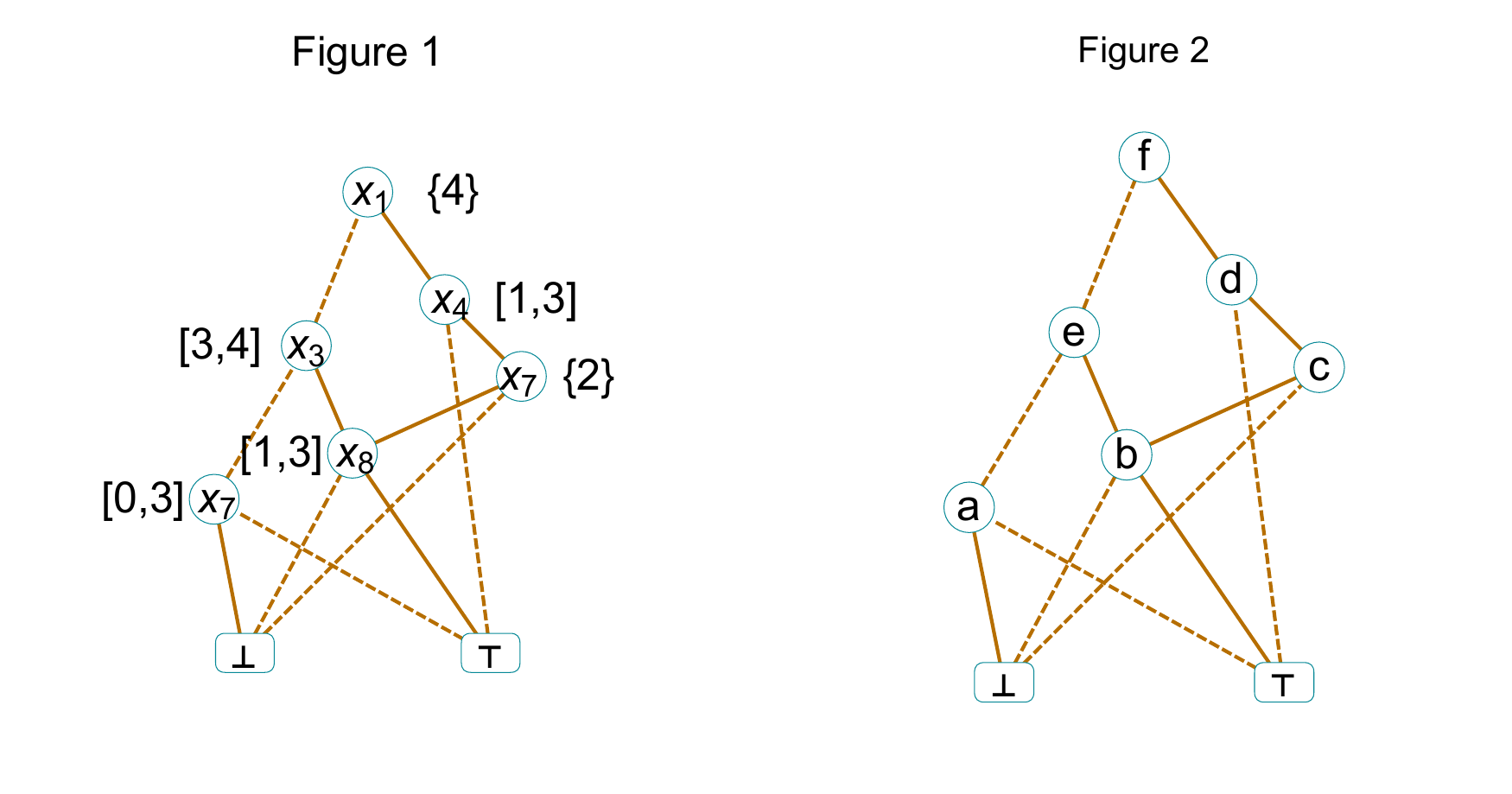}

For any Boolean function $\varphi = \varphi (x_1, \cdots, x_n)$ and any branching node $u$ of its\footnote{Upon fixing any ordering of the variables, each Boolean function has a unique BDD, see [K]. If in the sequel we speak of `the' BDD of $\varphi$ we silently assume that a suitable variable ordering has been fixed, and upon relabeling this ordering is $x_1, x_2,\ldots,x_n$.} BDD  we let $\mbox{ind}(u)$ be the index of the variable coupled to $u$. Furthermore it is handy to put $\mbox{ind}(\top) =\mbox{ind}(\perp) =n+1$. Most often our branching nodes have names among $a,b,c,d$, in which case we shall use $\alpha,\beta,\gamma,\delta$ as their corresponding indices.
Thus if $\varphi = \psi$ then $\alpha=\gamma= 7$ and $\mbox{ind}(\top) =11$. For any $\varphi$ and any branching node (say $c$) we denote by $\varphi_c$ the unique Boolean function defined by the induced BDD  with root $c$. Thus $\varphi_c = \varphi_c (x_\gamma, x_{\gamma+1}, \ldots, x_n)$.

{\bf 2.1} Suppose in the BDD of $\varphi$ node $a$ has the $0$-son $b$ and the $1$-son $c$. A moment's thought confirms that (see Algorithm $C$ in [K, p. 207]):

(1) \qquad $|\mbox{Mod}(\varphi_a)| = 2^{\beta-\alpha-1}\cdot |\mbox{Mod}(\varphi_b)|\  +\  2^{\gamma-\alpha-1}\cdot |\mbox{Mod}(\varphi_c)|$.

Using that $|\mbox{Mod}(\varphi_\perp)|=0$ and $|\mbox{Mod}(\varphi_\top)|=1$ this enables a quick recursive calculation of $|\mbox{Mod}(\varphi)|$. For $\varphi:=\psi$ we get (see Figures 1 and 2)

 \qquad $|\mbox{Mod}(\psi_a)|=2^{11-7-1}\cdot |\Mod(\psi_\perp)|+2^{11-7-1}\cdot |\Mod(\psi_\top)|=0+8\cdot 1=8$.

Upon shelling the BDD from below and repeatedly applying (1) one arrives at

(2)\qquad   $|\mbox{Mod}(\psi)| =|\mbox{Mod}(\psi_f)|  = 2^{3-1-1}\cdot 128   + 2^{4-1-1} \cdot 80 =576$.

{\bf 2.2} As opposed to  $|\mbox{Mod}(\varphi)|$, it is lesser known how to get the cardinalities $N_k:  = |\mbox{Mod}(\varphi, k)|$.  We present the method of [K, Exercise 25, p.260] with slightly trimmed notation. The unknown values $N_k$ get packed in a generating function

(3) \qquad $G(z) = G(z, \varphi) : = \ds\sum_{k=0}^n N_kz^k$.

For all branching nodes $a$ put $G_a (z) : = G(z, \varphi_a)$, as well as $G_\perp(z): = 0$  and $G_\top (z) : = 1$. Let $a,b,c$ be such that $b$ and $c$ are the $0$-son and $1$-son of $a$ respectively (either one of $b,\ c$ can be $\perp$ or $ \top$). Then

(4) \qquad $G_a (z) = (1+z)^ {\beta - \alpha -1} G_b(z) + z(1+z)^{\gamma - \alpha-1} G_c(z)$,

which resembles (1). In fact a recursive calculation akin to 2.1 yields

(5) \qquad $G(z)=1+ 8z + 30z^2 + 70z^3+113z^4 + 132z^5 + 113z^6 + 70z^7 + 30z^8 + 8z^9 + z^{10}$.

The coefficients add up to $576$ which matches (2). It is irrelevant that  they happen to be symmetric.

\section{Enumerating   Mod$(\varphi)$ and  Mod$(\varphi, k)$, the latter one-by-one }

The two kinds of model sets in the title are dealt with in Subsections 3.1 and 3.2 respectively.

{\bf 3.1} Let us review how a shelling of the BDD of $\varphi$ and the use of don't-care symbols provide a compressed enumeration of $\Mod(\varphi)$. Specifically, consider $\varphi:=\psi$ and suppose a shelling of $B_1$ in Fig.1 (see also Fig.2) so far yielded the compressed representations

$Mod(\psi_d)= (0,2,2,2,2,2,2)\uplus (1,2,2,1,1,2,2)$

$Mod(\psi_e)=(0,2,2,2,0,2,2,2)\uplus (1,2,2,2,2,1,2,2) $

where $\uplus$ means {\it disjoint union}. Furthermore, we prefer `2' over the usual don't-care symbol `$*$'. It is easy to verify the above formulas ad hoc. For instance, setting the root of $\psi_d$ to $x_4:={\bf 1}$ brings us to $x_7$, and obviously only $x_7=x_8:=1$ brings us to $\top$. This explains the {\it 012-row} $({\bf 1},2,2,1,1,2,2)$. Consider any $x\in\{0,1\}^{10}$. If $x$ has $x_1=0$ then $x\in \Mod(\psi_f)$ iff $(x_3,\ldots,x_{10})\in \Mod(\psi_e)$. 
 If $x$ has $x_1=1$ then $x\in \Mod(\psi_f)$ iff $(x_4,\ldots,x_{10})\in \Mod(\psi_d)$. From this it is clear that

(6) \quad Mod$(\psi) = (0,2,\ 0,2,2,2,0,2,2,2) \ \uplus \ (0,2,\ 1,2,2,2,2,1,2,2)$\\
 \hspace*{2.2cm} $\uplus \ (1,2,2\ ,0,2,2,2,2,2,2) \ \uplus  \ (1,2,2,\ 1,2,2,1,1,2,2)$.
 
{\bf 3.2} As opposed to calculating the {\it cardinality} of $\Mod(\varphi,k)$ (see 2.2), the literature seems to lack  a clever {\it enumeration} of $\Mod(\varphi,k)$. 

{\bf 3.2.1} The obvious naive way goes like this. Look at $\varphi:=\psi$ and the compressed enumeration of $\Mod(\psi)$ in (6). For say $k:=3$ the ${7}\choose{3}$ many 3-models contained in $(0,2,0,2,2,2,0,2,2,2)$ are easily enumerated. Similarly one proceeds for the 3-models in the second and third 012-row. However, the fourth 012-row contains {\it no} 3-models. This {\it empty-row-issue} prevents the naive method from running in polynomial total time.

{\bf 3.2.2} A more sophisticated method  was kindly pointed out to me by Fabio Somenzi. Let $B_1$ be the BDD in Figure 1 which thus triggers an enumeration of  Mod$(\varphi)$ as in 3.1. Construct a second BDD $B_2$ whose models are exactly the $k$-ones bitstrings in $\{0,1\}^n$. This is straightforward  and costs $O(nk)=O(n^2)$. Figure 3 shows $B_2$ when $n=10$ and $ k=4$.

\begin{center}
\includegraphics[scale=0.9]{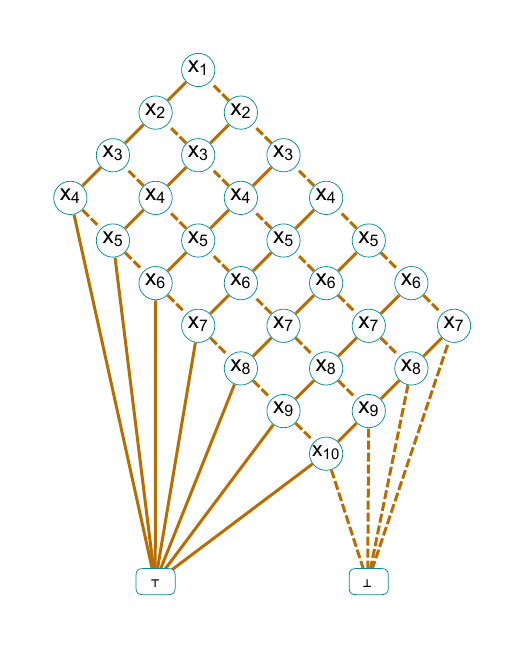}
\end{center}
{\sl Figure 3: This BDD $B_2$ accepts exactly the bitstrings $x\in\{0,1\}^{10}$ with $H(x)=4$.  }

Building the conjunction $B_3$ of $B_1$ and $B_2$ has polynomial compexity, and evidently the model set of $B_3$ equals Mod$(\varphi,k)$. This seems like a crisp polynomial total time procedure to the $k$-models of $\varphi$. Trouble is, $B_3$ may be much larger than $B_1$ (despite the adjective `polynomial'). Furthermore,
proceeding as in 3.1 to enumerate the models of $B_3$ necessarily deteriorates to a tedious one-by-one
 enumeration because no proper $012$-row can possibly consist entirely of $k$-models. To witness, for $\varphi:=\psi$ the diagram of $B_3$ is rendered\footnote{The notation $a1$ to $a10$ instead of $x_1$ to $x_{10}$ is due to technicalities of Python which was used to calculate the BDD. The first author, who uses exclusively Mathematica, is grateful to Jaco Geldenhuys for helping out with Python before the second author took over (in Section 6).} in Figure 4 below. Thus exactly 113 bitstrings in $\{0,1\}^{10}$ trigger $\top$, in Figure 4, in accordance with the summand $113z^4$ in (5).

\begin{center}
\includegraphics[scale=0.57]{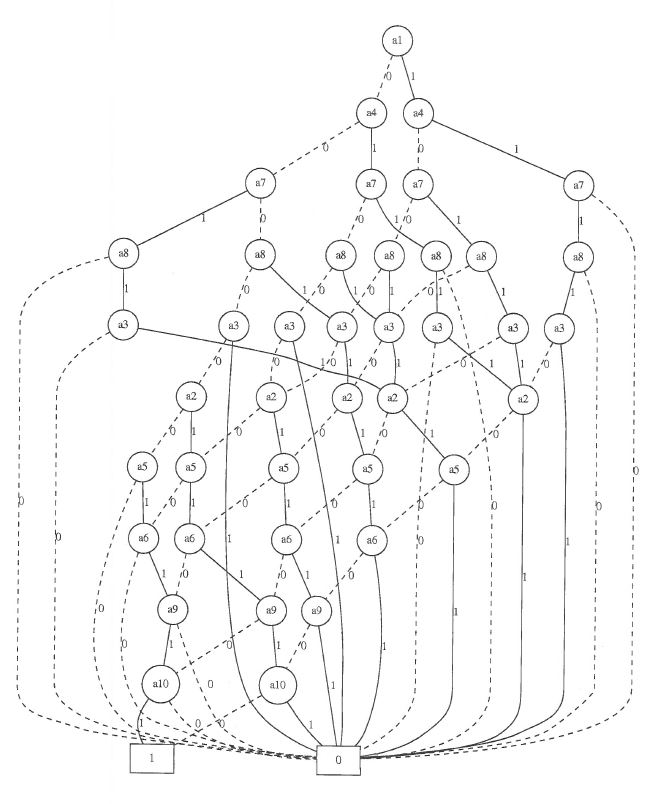}
\end{center}

{\sl Figure 4: The conjunction $B_3$ of $B_1$ and $B_2$.}

 Viewing that the BDD $B_2$ has a very symmetric structure one reviewer suggested to circumvent the full-blown intersection algorithm in order to calculate $B_3$ faster. (Recall that $B_3$ itself cannot shrink unless the variable ordering is changed.)
This might be possible, yet it remains the unpleasant fact that $B_3$ can't help to generate the $k$-models one-by-one (provided the standard method of 3.1 is used).

\section{Enumerating Mod$(\varphi,k)$ in a compressed format}

We refer to the introduction for an overview of Subsections 4.1 to 4.3.

{\bf 4.1} The crucial concept in the remainder of the article will be the wildcard
 $(g_t, g_t, \cdots, g_t)$ which means ``exactly $t$ digits 1 in this area''. Here $t \geq 1$ and the number of symbols $g_t$ must be strictly larger than $t$. That's because $(g_4, g_4, g_4)$ is impossible, and instead of $(g_3, g_3, g_3)$ we stick to $(1, 1, 1)$. As an application consider the compressed representation of $\Mod(\psi)$ in (6). It immediately triggers the representation

(7) \qquad Mod$(\psi, 4) = (0, g_4, 0, g_4, g_4, g_4, 0, g_4, g_4, g_4) \uplus (0, g_2, 1, g_2, g_2, g_2, g_2, 1, g_2, g_2)$

\hspace*{1.5cm} $\uplus\, (1, g_3, g_3, 0, g_3, g_3, g_3, g_3, g_3, g_3) \uplus (1, 0, 0, 1, 0, 0, 1, 1, 0, 0)$

Condensing Mod$(\varphi, k)$ in this way worked well here but imagine  that $r \cap \mbox{Mod}(\varphi, k) = \emptyset$ for 99 million out of 100 million rows $r$. Thus we run again into the empty-row-issue\footnote{Our naive way of 3.2.1 (and with it the empty-row-issue) is not bound to BDD's. It can be applied whenever the modelset of a Boolean function is displayed as a disjoint union of 012-rows. This can e.g. be achieved by the Mathematica command {\tt BooleanConvert}; see also [W,Sec. 6.1].} of 3.2.1. It will be tackled in 4.2.

{\bf 4.2} The following notations are handy. For $\varphi=\varphi(x_1,\ldots,x_n)$ and $S\subseteq [0,n]$  put $Mod(\varphi,S):=\bigcup_{i\in S} Mod(\varphi,i)$.
For integers $0 \leq u <  v$  put $[u,v] : = \{u, u+1, \cdots, v\}$.
 In order to glimpse how polynomial total time will be achieved we invoke again $\varphi:=\psi$. For the time being, esteemed reader,  blindly accept that the 01g-rows in Tables 1 and 2 provide compact representations of $\Mod(\psi_d,[1,3])$ and $\Mod(\psi_e,[3,4])$ respectively. (Specifically, e.g. the disjoint union of the last two 01g-rows in Table 1 is $Mod(\psi_d,3)$.) This enables one to build a compact representation of
$\Mod(\psi,4)$ akin to how $\Mod(\psi)$ arose from $\Mod(\psi_d)$ and $\Mod(\psi_e)$ in 3.1. For instance the set $(1,1,1)\times Mod(\psi_d,1)$ of all concatenations of $(1,1,1)$ with bitstrings from $Mod(\psi_d,1)$ is a subset of $Mod(\psi,4)$. Similarly $(1,g_1,g_1)\times Mod(\psi_d,2)$ and $(1,0,0)\times Mod(\psi_d,3)$ are subsets of  $Mod(\psi,4)$. Turning from $\psi_d$ to $\psi_e$, likewise  $(0,1)\times Mod(\psi_e,3)$ and $(0,0)\times Mod(\psi_e,4)$ are subsets of $Mod(\psi,4)$. Arguing similarly as in 3.1 the union of all these subsets actually {\it exhausts} $Mod(\psi,4)$.

\begin{tabular}{|c|c|c|c|c|c|c|c|} 
$x_4$ & $x_5$ & $x_6$ & $x_7$ & $x_8$ & $x_9$ & $x_{10}$ & Hamming-weight \\ \hline
     &       &        &       &       &       &        & \\ 
 0 & $g_1$ & $g_1$ & $g_1$ & $g_1$ & $g_1$ & $g_1$ & 1 \\
 0 & $g_2$ & $g_2$ & $g_2$ & $g_2$ & $g_2$ & $g_2$ & 2 \\
 0 & $g_3$ & $g_3$ & $g_3$ & $g_3$ & $g_3$ & $g_3$ & 3 \\
 1 & 0 & 0 & 1 & 1 & 0 & 0 &  3\\
\hline
 \end{tabular}
 
 {\sl Table 1: The 01g-rows of of Mod$(\psi_d, [1,3])$, sorted by their Hamming-weights}

\vspace{1cm}

\begin{tabular}{|c|c|c|c|c|c|c|c|c|} 
$x_3$ & $x_4$ & $x_5$ & $x_6$ & $x_7$ & $x_8$ & $x_9$ & $x_{10}$ & Hamming-weight \\ \hline
 &          &         &      &      &      &       &     & \\ \hline 
 0 & 0 & 0 & 0 & 0 & 1 & 1 & 1 & 3 \\ 
 0 & $g_1$ & $g_1$ & $g_1$ & 0 & $g_2$ & $g_2$ & $g_2$ & 3 \\
 0 & $g_2$ & $g_2$ & $g_2$ & 0 & $g_1$ & $g_1$ & $g_1$ & 3 \\
 0 & 1 & 1 & 1 & 0 & 0 & 0 & 0 & 3 \\
  1 & 0 & 0 & 0 & 0 & 1 & $g_1$ & $g_1$ & 3\\
 1 & $g_1$ & $g_1$ & $g_1$ & $g_1$ & 1 & 0 & 0 & 3 \\
 0 & $g_1$ & $g_1$ & $g_1$ & 0 & 1 & 1 & 1 & 4 \\
 0 & $g_2$ & $g_2$ & $g_2$ & 0 & $\ol{g}_2$ & $\ol{g}_2$ & $\ol{g}_2$ & 4 \\
 0 & 1 & 1 & 1 & 0 & $g_1$ & $g_1$ & $g_1$ & 4\\
 1 & 0 & 0 & 0 & 0 & 1 & 1 & 1 & 4\\
 1 & $g_1$ & $g_1$ & $g_1$ & $g_1$ & 1& $\ol{g}_1$ & $\ol{g}_1$ & 4\\
 1 & $g_2$ & $g_2$ & $g_2$ & $g_2$ &1 & 0 & 0 & 4 \\ \hline
 \end{tabular}

{\sl Table 2: The 01g-rows of of Mod$(\psi_e, [3,4])$, sorted by their Hamming-weights}

{\bf 4.3} Let the BDD $B$  of $\varphi=\varphi(x_1,\ldots,x_n)$ be known and fix $k\in [0,n]$. For each branching node $c\in B$ let $Touch(c)$ be the set of all $x\in Mod(\varphi,k)$ whose accepting path contains the node $c$. The {\it triggered Hamming-weight set} is defined as

$THS(c):=\{H(x_\gamma, x_{\gamma+1},\ldots, x_n):\ x\in Touch(c) \}$.

For instance, since $Mod(\varphi,k)$ will be assumed to be nonvoid, the root $x_1$ has $THS(x_1)=\{k\}$. The {\it relevant set}  

${\cal R}:=\{a\in B\setminus\{\perp,\top\}:\ THS(a)\neq \emptyset\}$

has the property that each $a\in {\cal R}$ has at least one son in $ {\cal R}\cup\{\top\}$.
For instance,  for $\varphi:=\psi$ and $k:=1$ one has  $ {\cal R}=\{a,d,e,f\}$.  If  $k:=4$ then it follows from Table 1 that $THS(d)=[1,3]$, and from Table 2 that $THS(e)=[3,4]$. Generally the sets next to the branching nodes of Figure 1 are the corresponding THS-sets.
We postpone the proof of claim (8) to Section 5.

\begin{itemize}
\item[(8)]   For the BDD $B$ of any Boolean function $\varphi$ with $Mod(\varphi,k)\neq \emptyset$ the triggered Hamming-weight sets $THS(a)$ can be calculated in time $O(|B|n^2)$.
\end{itemize}

{\bf Theorem:} {\it There is an algorithm which achieves the following. Given the BDD $B$ of a Boolean function $\varphi=\varphi(x_1,\ldots,x_n)$ and given $k\in [0,n]$, it renders $\Mod(\varphi,k)$ as a union of $N$ disjoint 01g-rows in time $O(|B|\cdot n\cdot max(n,N)))$.}

{\it Proof.} Using the method of 2.2 one can predict the cardinality $|\Mod(\varphi,k)|$ in time $O(|B|n)$.  If $\Mod(\varphi,k)=\emptyset$, we are done. Otherwise all sets $THS(a)$ can be calculated in time $O(|B|n^2)$ by (8). Shelling ${\cal R}$ from below we now  prove by induction that each  $\Mod(\varphi_a, THS(a))\ (a\in {\cal R})$ can be written\footnote{Of course $a\in {\cal R}$ concerns an arbitrary BDD, i.e. $a$ is not the $a$ in Figure 2.}
as disjoint union of 01g-rows. 

Specifically, we will embark on a 4-fold case distinction concerning the kinds of sons of $a$. Namely its 0-son $b$ either satisfies $b\in {\cal R}$ or $b\not\in {\cal R}$  or $b=\top$. (Note that $b=\perp$ is a special case of $b\not\in {\cal R}$.) Similarly the 1-son $c$ has exactly one of the three properties. Since it will be evident that e.g. the scenario ($b\not\in {\cal R},c\in {\cal R}$) is handled dually to
($c\not\in {\cal R},b\in {\cal R}$), we can capture both by the acronym $\{\in {\cal R},\not\in {\cal R}\}$ (this is Case 4 below). The three other cases are $\{\in {\cal R},\in {\cal R}\}$ (Case 2) and $\{\in {\cal R},\top\}$ (Case 3) and $\{\not\in {\cal R},\top\}$ (Case 1). Observe that $\{\top,\top\}$ is impossible, and so is $\{\not\in {\cal R},\not\in {\cal R}\}$ (why?). We begin with the easiest case.

{\it Case 1:} Let  $a\in {\cal R}$ be such that one son is  $\top$ and the other is $\not\in {\cal R}$. If say
$c\not\in {\cal R}$ then $Touch(c)=\emptyset$. Since $c$  cannot contribute to $\Mod(\varphi_a, THS(a))$, the task is on $b=\top$. To fix ideas, say $THS(a)=[2,4]$ and $\alpha=n-4$. Since $b=\top$ is the 0-son, we conclude

(9)\quad $\mbox{Mod}(\varphi_a,THS(a))=\mbox{Mod}(\varphi_a,2)\uplus \mbox{Mod}(\varphi_a,3)\uplus \mbox{Mod}(\varphi_a,4)$

$\hspace{4cm} =(0,g_2,g_2,g_2,g_2)\uplus (0,g_3,g_3,g_3,g_3)\uplus(0,1,1,1,1)$.

If instead the 1-son $c$ equals $\top$, then (still assuming $\alpha=n-4$) we conclude

(10)\quad  $\mbox{Mod}(\varphi_a,THS(a))=(1,g_1,g_1,g_1,g_1)\uplus (1,g_2,g_2,g_2,g_2)\uplus (1,g_3,g_3,g_3,g_3)$.

{\it Case 2:} Both $b,c\in {\cal R}$. By the definition of $THS(a)$ for each fixed $j\in THS(a)$ there is at least one bitstring $(x_\alpha,x_{\alpha+1},\ldots,x_n)\in \Mod(\varphi_a)$ of Hamming-weight $j$. When fed to $B_a$ its accepting path must trace either $b$ or $c$. Hence at least one of two kinds of contribution towards the compact representation of $\Mod(\varphi_a,j)$ takes place.

Namely, whenever $b$ gets traced, one has $(x_\alpha,\ldots,x_n)=({\bf 0},?,...,?,x_\beta,...,x_n)$. Let $i\ge 0$ be the number of $1$-bits among $?,...,?$. Call any $i$ arising this way $(b,j)$-{\it admissible}. Because $j-i\in THS(b)$, by induction $\Mod(\varphi_b,j-i)$, which is part of $\Mod(\varphi_b,THS(b))$, has been written as disjoint union of 01g-rows.
 It follows that

 $S(j,i):=({\bf 0},g_i,\ldots,g_i)\times \Mod(\varphi_b,j-i)$

is a subset of $Mod(\varphi_a,j)$ for each $(b,j)$-admissible $i$. If $\sigma(j,i)$ is the number of 01g-rows constituting $Mod(\varphi_b,j-i)$ then also $\sigma(j,i)$ rows constitute 
$S(j,i)$. Because distinct $(b,j)$-admissible $i,\ i'$ induce disjoint subsets of $Mod(\varphi_b,j-i)$ and $Mod(\varphi_b,j-i')$ of $Mod(\varphi_b, THS(b))$, the number $\sigma(j)$ of 01g-rows triggered by one fixed $j\in THS(a)$ is bounded by the number of 01g-rows constituting $Mod(\varphi_b, THS(b)$.

Letting $j$ range over $THS(a)$ we obtain the set $S$ of all $x\in Mod(\varphi_a,THS(a))$ whose accepting path traces $b$ as a disjoint union of at most
$|THS(a)|\cdot |Mod(\varphi_b,THS(b))|$ many 01g-rows.
Likewise the set $T$ of all bitstrings $x\in Mod(\varphi_a,THS(a))$ whose accepting path traces $c$ is made up by chunks of type

$T(j,i):=({\bf 1},g_i,\ldots,g_i)\times \Mod(\varphi_c,j-(i+1)),$

and as above one concludes that $T$ can be written as a disjoint union of at most $|THS(a)|\cdot |Mod(\varphi_c,THS(c))|$ many 01g-rows.
 Similar to 3.1 and 4.2 one argues that the subset $S\cup T$ actually {\it exhausts}  $ \Mod(\varphi_a,THS(a))$.

{\it Case 3:} Only one of $b,c$ belongs to $ {\cal R}$, the other is $\top$. Then one of the two kinds of contributions in Case 2 gets replaced by  a type (9) or type (10) contribution occuring in Case 1.

{\it Case 4:} Only one of $b,c$ belongs to $ {\cal R}$, the other is {\it not} in $ {\cal R}$. Then one of the two kinds of contributions in Case 2 evaporates altogether.

As to the cost analysis, for each $ \Mod(\varphi_a,THS(a))$ the number of 01g-rows $r$ in its compressed representation is at most $N$ because $r$ occurs as suffix in at least one 01g-row constituting $ \Mod(\varphi_a,k)$. Since these 01g-rows $r$ are (readily obtained) concatenations of bitstrings of length at most $n$, the cost of constructing any individual $ \Mod(\varphi_a,THS(a))$ is $O(Nn)$. The overall cost of our algorithm therefore is

$O(|B|n^2)+|B| O(Nn)=O(|B|\ n\ max(n,N)).\ \square$

{\bf 4.4} We define the {\it height} $h(B)$ of the BDD $B$ of $\varphi(x_1,\ldots,x_n)$ as the length of a longest chain from the root to $\top$. Furthermore, the {\it bottom-gap} $bot(B)$ is the maximum of the values $n - ind(c)$ where $c$ ranges over $min({\cal R})$. Furthermore, let
 $\kappa(B):=max\{|THS(a)|\! :\! a\in {\cal R}\setminus min({\cal R}\}$. From
$THS(a)\subseteq [0,k]$ follows $\kappa\le k+1$. For instance, the BDD $B_1$ in Figure 1 has $\kappa(B_1)=|[0,3]|=4,\ h(B_1)=4$, and $bot(B)=10-7=3$.

{\bf Corollary:} {\it Let $\varphi:\{0,1\}^n\to\{0,1\}$ have a BDD $B$ with parameters $h:=h(B),\ bot:=bot(B),\ \kappa:=\kappa(B)$. Then for any $k\in [0,n]$ the set $Mod(\varphi,k)$ can be enumerated as a disjoint union of at most $bot\cdot (2\kappa)^{h-1}$ many 01g-rows.}

{\it Proof.} Using any shelling of ${\cal R}$ it suffices to show by induction that all $Mod(\varphi_a,THS(a))\ (a\in {\cal R})$ can be enumerated with at most
$bot\cdot (2\kappa)^{h(B_a)-1}$ many 01g-rows. To begin with, this holds for all minimal branching nodes $a$ in view of

$bot\cdot (2\kappa)^{h(B_a)-1}=bot\cdot (2\kappa)^0=bot.$

Suppose now $a$ has sons $b$ and $c$. By induction  $Mod(\varphi_b,THS(b))$ and $Mod(\varphi_c,THS(c))$ can be represented by at most $bot\cdot (2\kappa)^{h(B_b)-1}$
and $bot\cdot (2\kappa)^{h(B_c)-1}$ rows respectively.
Without loss of generality we can assume that $h(B_b)\le h(B_c)$. Then clearly $h(B_a)= h(B_c)+1$. Looking at the proof of the Theorem it follows that $Mod(\varphi_a,THS(a))$
can be represented by at most

$\kappa(bot\cdot  (2\kappa)^{h(B_b)-1}) + \kappa(bot\cdot  (2\kappa)^{h(B_c)-1}) \le 2\kappa(bot\cdot  (2\kappa)^{h(B_c)-1}=bot\cdot(2\kappa)^{h(B_c)} =bot\cdot(2\kappa)^{h(B_a)-1} $

many 01g-rows. $\square$

\section{How to calculate the sets THS(a)} 

The first step (in 5.1) in our quest to calculate all sets THS($a$) for arbitrary Boolean functions $\varphi$ is to determine some obvious supersets
 $THS'(a)\supseteq THS(a)$. In 5.2  this inclusion gets improved to $THS(a)\subseteq  THS'(a)\cap THS^*(a)$ for some suitably defined set 
$THS^*(a)$. In 5.2 we establish that actually $\subseteq$ is $=$, from which the pending proof of (8) ensues at once.

 {\bf 5.1} As opposed to previous shellings here we shell the BDD {\it  from above} in order to recursively define $THS'(a)$. For starters,  $THS'(x_1):=THS(x_1)=\{k\}$. By induction assume that for the upper neighbours $c,d,\ldots$ of node $b$ the sets $THD'(c),\  THD'(d),...$   have been calculated. Then $THD'(b)$ by definition contains two types of numbers $j$. Namely, for all upper neighbours $d$ to which $b$ is a $0$-son, do the following. For each $i$ in  
$THD'(d) $ the non-negative ones among these numbers become type-1 numbers $j$ in $THD'(b)$:

\qquad $i,\ i-1,\ldots,i-(\beta-\delta-1)$

Similarly, for all upper neighbours $c$ to which $b$ is a $1$-son, do the following. For each $i\in THD'(c)$ the non-negative ones among these numbers become
 type-2 numbers $j$ in $THD'(b)$:

$i-1,\ (i-1)-1,\ldots,(i-1)-(\beta-\gamma-1)$

In other words,  $j=i-t$ where $t$ ranges over $ [1,\beta-\gamma]\cap[0,i]$.

\includegraphics[scale=0.47]{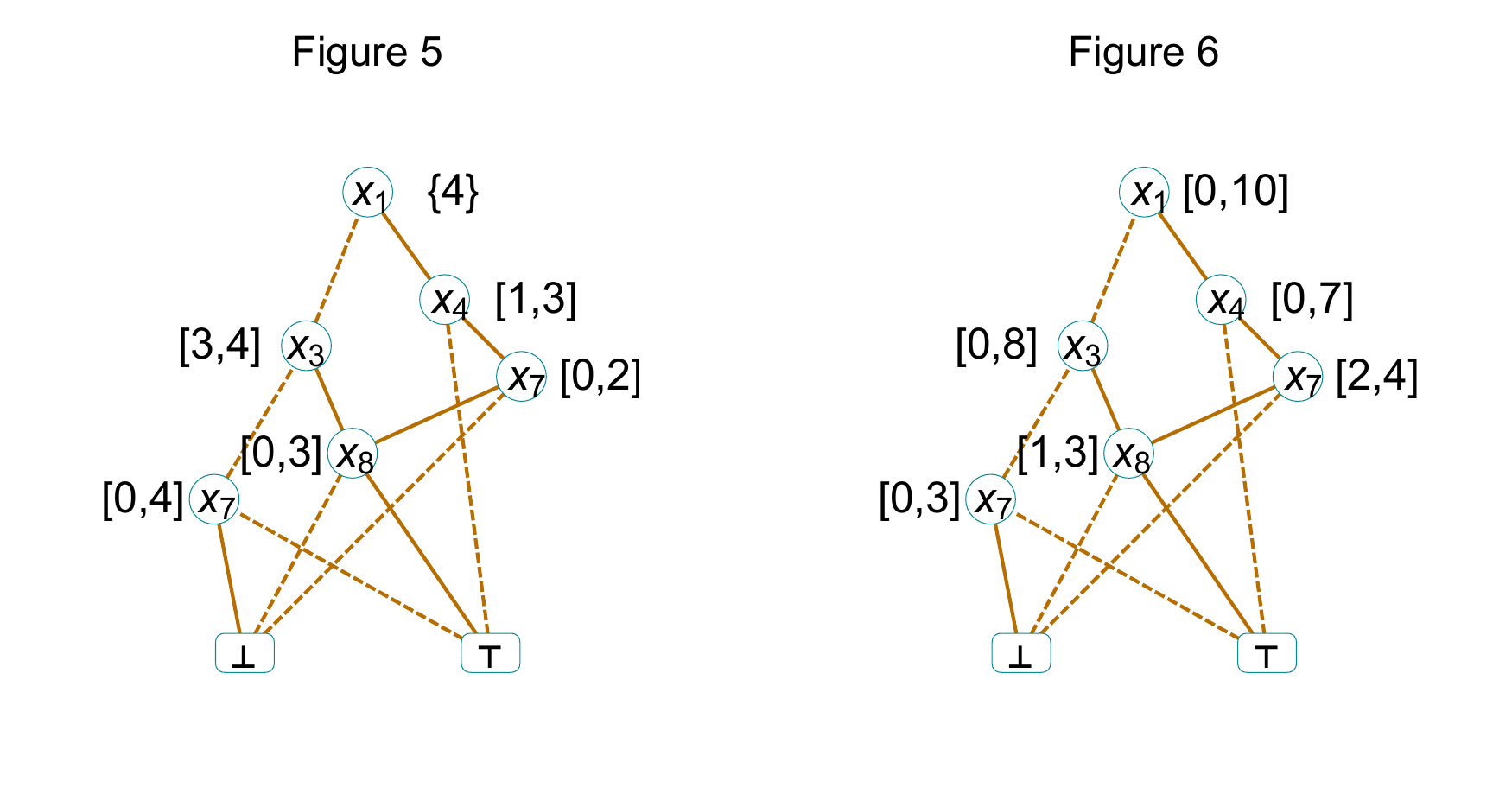}

In Figure 5 each node in the BDD of $\psi$ is labeled by its THS'-set. For instance,  $b=x_8$ (see Fig.2) has the upper neighbours $c$ and $e$. Because $b$ is the $1$-son of both $c$ and $e$, the set $THS'(b)$ receives from  $THS'(e)=[3,4]$ the numbers $3-1,\ 3-2,\ 3-3$ and $4-1,\ 4-2,\ 4-3,\ 4-4$. Likewise $THS'(b)$ receives from  $THS'(c)=[0,2]$ the numbers $1-1$ and $2-1$. Therefore $THS'(b)=[0,3]$.

{\bf Lemma 1:} {\it Suppose $\varphi=\varphi(x_1,\ldots,x_n)$ is such that $\Mod(\varphi,k)\not=\emptyset$. Let $a\neq x_1$ be a branching node in the BDD of $\varphi$ which has $THS'(a)\neq\emptyset$. Then for any $j\in THS'(a)$ there is an upper neighbour $b$ of $a$ and a bitstring $y'=(y_1,y_2,...,y_\beta,...,y_{\alpha-1})$ with these properties:

(i) Feeding $(y_1,y_2,...,y_\beta)$ to the BDD triggers a path from the root to $a$;

(ii)  $H(y')+j=k$. }

For instance, take $THS'(a)=[0,4]$ in Figure 5 and let $j:=1$. Since $a$ has only one upper neighbour $e$, it is up to $e$ providing the bitstring $y'$ postulated in Lemma 1. Indeed, e.g. set $y'=(y_1,y_2,y_\epsilon,y_4,y_5,y_{\alpha-1}):=(0,1,0,1,1,0)$. Property (i) holds since feeding $(0,1,0)$ to the BDD in Figure 5 brings us from the root to $a$. And (ii) holds since $H(y')+j=3+1=k$.

{\it Proof of Lemma 1.} We induct on the longest length of a path from the root to $a$. As anchor, let $a$ be a son of the root $b\ (=x_1)$, say the $0$-son (the $1$-son is treated similarly). Pick any $j\in THS'(a)$.  Then $j=k-t$ for some $t\in [0,\alpha-\beta-1]=[0,\alpha-2]$ by the very definition of $THS'(a)$. Put $y':=(y_1,y_2,...,y_{\alpha-1})=(y_\beta,y_2,...,y_{\alpha-1})$, where $y_\beta:= {\bf 0}$ and exactly (any) $t$ bits among
$y_2,...,y_{\alpha-1}$ must be $1$, the others $0$. Then (i) holds since $a$ is the $0$-son of $b$ and $y_\beta:= 0$. And (ii) holds since
$H(y')+j=t+(k-t)=k$.

If $a$ is not a son of the root and $j\in THS'(a)$ then by the recursive definition of the $THS'$-sets there is some upper neighbour $b$ of $a$ such that some $i\in THS'(b)$ gave rise to $j$. Specifically, if say $a$ is the $1$-son of $b$ then, recall, $j=i-t$ for some $t\in [1,\alpha-\beta]$. By induction there is a bitstring $(y_1,...,y_{\beta-1})$ of Hamming-weight $k-i$ which when fed (up to its last few bits) to the BDD brings us from $x_1$ to $b$. Consider the bitstring

$y':=(y_1,...,y_{\beta-1},{\bf 1},1,...,1,0,...,0)$

of length $\alpha-1$ which has exactly $t$ many $1$'s after $y_{\beta-1}$ (and thus $(\alpha-\beta)-t\ge 0$ many $0$'s). Because feeding 
$(y_1,...,y_{\beta-1},1)$ to the BDD brings us from $x_1$ to $a$, property (i) holds. Also (ii) holds since $H(y')+j=  ((k-i)+t)+j=k.\ \square$

{\bf 5.2} Recall that always $THS'(a)\supseteq THS(a)$. For instance $THS'(c)=[0,2]$ in Figure 5 is a strict superset of  $THS(c)=\{2\}$ in Figure 1.
In order to zoom in from $THS'(a)$ to $ THS(a)$ for a general $\varphi$ we put

(11)\qquad $THS^*(a):=\{i\in [0,n]:\ \Mod(\varphi_a,i)\neq\emptyset\}$.

While $THS(a)\subseteq  THS'(a)\cap THS^*(a)$ is evident, the converse inclusion is not.

{\bf Lemma 2:} {\it Suppose $\varphi=\varphi(x_1,\ldots,x_n)$ is such that $\Mod(\varphi,k)\not=\emptyset$. Then for all branching nodes $a$ in the BDD 
of $\varphi$ it holds that $THS(a)=THS'(a)\cap THS^*(a)$.}

{\it Proof.} Showing that $j\in THS'(a)\cap THS^*(a)$ implies $j\in THS(a)$ amounts to exhibit a bitstring $y=(y_1,...,y_\alpha,...,y_n)\in Touch(a)$ with
 $H(y_\alpha,...,y_n)=j$. First, since $j\in THS'(a)$, there is a bitstring $y'=(y_1,...,y_{\alpha-1})$ with properties (i) and (ii) of Lemma 1. Second, since 
$j\in THS^*(a)$, there is a bitstring $y^*=(y_{\alpha},...,y_n)\in \Mod(\varphi_a,j)$. Define $y$ as the concatenation of $y^*$ and $y'$. Since $y'$ satisfies (i) and $y^*\in \Mod(\varphi_a,j)$ we conclude that $y\in \Mod(\varphi)$, that $a$ is in the accepting path of $y$, and that $H(y^*)=j$. Finally, since $y'$ satisfies (ii), we have $H(y')=k-j$, hence $H(y)=H(y')+H(y^*)=k$, hence $y\in\Mod(\varphi,k)$, hence $y\in Touch(a).\ \square$.

{\bf 5.2.1} In order to calculate $THS^*(a)$ we  shell the BDD from below. For starters
 $THS^*(\perp)=\emptyset$ and $THS^*(\top)=\{0\}$. Put $j+S:=\{j+i:\ i\in S\}$. When $a$ has the $0$-son  and $1$-son $b$ and $c$ respectively, then evidently

(12) \quad  $THS^*(a) = \ds\bigcup_{j=0}^ {\beta-\alpha-1} \bigg(j + THS^*(b)\bigg)\ds  \cup \ds \bigg(1+\bigcup_{j=0}^{\gamma-\alpha-1} (j+ THS^*(c))\bigg)$

\quad  $\hspace{2.5cm} = \ds\bigcup_{j=0}^ {\beta-\alpha-1} \bigg(j + THS^*(b)\bigg) \cup \ds \bigcup_{j=1}^{\gamma-\alpha} \bigg(j+ THS^*(c)\bigg)$.

Applying (12) to the BDD of $\varphi:=\psi$ yields the $THS^*$-sets indicated in Figure 6. One checks that the intersection of corresponding intervals in Figures 5 and 6 indeed yields the right interval in Figure 1; say $[0,2]\cap [2,4]=\{2\}$.

{\bf 5.3}  As to the cost of calculating all $THS'(b)$, by 5.1 finding the contribution of an upper neighbour $d$ of $b$ towards $THS'(b)$ costs $O(n^2)$. If $u(b)$ is the number of upper neighbours of $b$ then the sum of all $u(b)$'s is slightly\footnote{The lines incident with $\perp$ or $\top$ do not occur.} less than the number of lines in $B$, which in turn is less than $2|B|$. It follows that
calculating all $THS'(b)$ costs $O(|B| n^2)$.

Since $\beta-\alpha-1$ and $\gamma-\alpha-1$ in (12) are bound by $n$, it readily follows that calculating all sets $THS^*(a)$ also costs  $O(|B| n^2)$. 
From the above and $THS(a)=THS'(a)\cap THS^*(a)$ (Lemma 2) follows that the cost of calculating all sets $THS(a)$ is 
$O(|B| n^2)+O(|B| n^2)+O(|B| n)=O(|B| n^2)$,
which proves claim (8).

\section{Numerics}

Recall that a {\it clause} of {\it length} $q$ is a Boolean formula that is a conjunction of $q\ge 1$ different literals. A {\it conjunctive normal form (CNF)} is a Boolean formula that is a conjunction of clauses. For values of $p,q,n$ we generate Boolean functions $\varphi:\{0,1\}^n\to\{0,1\}$ which are given by a random CNF formula. Specifically, this CNF has $p$ clauses, all of which of length $q$. 
Using the Python command {\tt expr2bdd} a BDD of $\varphi$ was computed.
Using this BDD we fix some $k$ and calculate $Mod(\varphi,k)$ as a disjoint union of $01g$-rows and record $N:=|Mod(\varphi,k)|$. The $01g$-rows are either calculated in the naive or the sophisticated way. The corresponding numbers of $01g$-rows are $R1,\ R2$, and the corresponding CPU-times are $T1,\ T2$. Times $>10sec$ were rounded to full seconds. Furthermore $R1'$ gives the number of empty rows we incur when using the naive approach. 

Usually the number of empty rows is less than the number of non-empty rows  but there are  exceptions, e.g. in 6 of the 7 first rows of Table 3.  For small $n$ there is no significant difference between the CPU-times of both methods.
But for higher $n$ the sophisticated method excels; it is more than 100 times faster in  rows 8-13 of Table 3.  In rows 14-20 the CPU-times still differ significantly, but what springs to mind even more are the different degrees of compression.
For instance in row 19 the naive way required 113477 $01g$-rows to compress $Mod(\varphi,65)$, whereas the sophisticated way required only 964.

Finally, since the time complexity of the model is linear with respect to the size $|B|$ of the BDD, we assessed the sensitivity of $|B|$ with respect to the parameters $p$ and $q$. We found  that $|B|$ is more sensitive to increasing values of $p$ than it is for increasing values of $q$. For instance when $n, p, q$ are $40, 25$ and $5$ respectively, $|B|$  is approximately $100$ times higher that when $n, p,$ and $ q$ are $40, 5$ and $25$ respectively. 

\begin{center}
\begin{small}
\begin{tabular}{|c|c|c|c|c|c|c|c|c|c|c|} 
 & $n$ & $p$ & $q$ & $k$ & $N$ & $R1$ & $R1'$ & $T1$ & $R2$ &  $T2$ \\ \hline
 &    &       &        &       &       &       &     &  &    & \\ 
1& $10$  &  $7$ & $3$   & $3$ & $49$    & $18$  & $39$    & $0.0008$ & $16$ & $0.0005$  \\
2& $18$  & $16$ & $9$  & $5$ & $8'262$ & $365$ & $1'120$ & $0.02$   & $75$ & $0.007$ \\
3& $20$ &  $15$ & $11$  & $7$ & $76'417$ & $922$ & $1'006$ & $0.03$ &$393$ & $0.01$\\
4& $27$ & $16$ & $10$ & $21$ & $293'888$ & $4'127$ & $22'248$ & $0.51$ &$977$ &$0.05$ \\
5& $ 27$ & $20$ & $15$ & $10$ & $8'436'132$ & $6'350$ & $1'746$ & $0.27$ &$1'852$ &$0.08$ \\
6& $32$ & $25$ & $27$ &  $24$ & $10'518'300$ &      $592$ & $930$ & $0.02$ & $121$ & $0.01$\\
7& $50$ & $45$ & $33$ & $40$ & $ 10'272'281'712$ & $10'979$ & $26'312$ & $1.25$ & $9'836$ & $1.00$ \\ \hline
8& $67$ & $74$ & $36$ & $42$ & $1.7\times 10^{18}$ & $400'119$ & $25'741$ & $1'873$ & $ 12'374$ & $ 14$ \\
9& $74$ & $97$ &$39$ &$40$ & $1.4\times 10^{21}$ & $ 875'833$ & $32$ & $9'646$ & $245'837$ & $32$\\
10& $ 98$ & $132$ &$60$ & $53$ &$1.8\times 10^{28}$ & $ 550'113$ & $545$ & $4'479$ & $ 67'287$ & $30$\\
11&  $101$ & $72$ & $54$ & $42$ & $4.8\times 10^{28}$ & $916'032$ & $222$ & $16'137$ & $118'739$ & $60$\\
12& $103$ & $77$ & $54$ & $81$ & $1.5\times 10^{22}$ & $910'531$ & $736'193$ & $21'247$ & $358'314$ & $117$ \\
13& $110$ & $67$ & $54$ & $77$ & $9.6\times 10^{28}$ & $1'166'067$ & $83'161$ & $ 23'256$ & $ 324'746$ & $83$ \\ \hline
14& $28$ & $11$ & $8$ & $13$ & $50'645'009$ & $ 87'725$ & $20'911$ & $0.13$ & $ 7'582$ & $0.02$ \\
15& $78$ & $102$ & $54$ & $34$ & $ 1.4\times 10^{22}$ & $117'797$ & $2'276$ & $107$ & $9'729$ & $4.51$\\
16& $90$ & $153$ & $65$ & $39$ & $4.7\times 10^{25}$& $191'498$ & $1'627$ & $596$ & $2'759$ & $9.93$ \\
17& $92$ & $88$ & $75$ &  $5$ & $4.0\times 10^{26}$ & $ 31'571$ & $135$ & $6.12$ & $963$ & $1.37$ \\
18& $96$ & $89$ & $65$ & $52$ & $4.6\times 10^{27}$ & $ 143'508$ & $512$ &$ 239$ & $10'374$ & $ 6.88$ \\
19& $119$& $77$ & $81$ & $65$ & $2.93\times 10^{34}$ & $113'477$ & $763$ & $225$ & $964$ & $8.22$ \\
20& $120$ & $86$ & $91$ & $72$ & $8.8\times 10^{33}$ & $66'218$ & $1'883$ & $80$ & $903$ & $5.08$ \\
\hline
 \end{tabular}
\end{small}

{\sl Table 3: Numerical comparison of the naive way and the sophisticated way}

 \vspace{3cm}

\end{center}

{\bf Acknowledgment:} The first author thanks Fabio Somenzi and Moshe Vardi for fruitful comments.

\section*{References}
\begin{enumerate}

\item[{[ABJM]}]  Antoine Amarilli, Pierre Bourhis, Louis Jachiet, and Stefan Mengel, A Circuit-Based Approach to Efficient Enumeration", ICALP 2017
\item[{[CH]}] Y. Crama, P.L. Hammer, Boolean Functions, Enc. of Math. Appl. 142, Cambridge University Press 2011.
	\item [{[K]}] D.E. Knuth, The Art of Computer Programming, Volume 4A, Combinatorial Algorithms, Part 1, Addison Wesley 2012.
	\item[{[W]}] M. Wild, Compression with wildcards: Abstract simplicial complexes, Quaestiones Mathematicae 46 (2023) 1151-1173.

\end{enumerate}

\end{document}